\documentclass[mathleft]{an}
\usepackage{graphicx}
\usepackage{times}

\usepackage{natbib}
\bibpunct{(}{)}{;}{a}{}{,}
\usepackage{graphicx}
\usepackage{times}

\begin{document}

\Pagespan{789}{}
\Yearpublication{2006}%
\Yearsubmission{2005}%
\Month{11}%
\Volume{999}%
\Issue{88}%

\title{Oscillator Models of the Solar Cycle and the Waldmeier Effect}

\author{M. Nagy
\and  K. Petrovay
}
\titlerunning{}
\authorrunning{}
\institute{
E\"otv\"os University, Department of Astronomy, Budapest, Hungary}

\received{30 May 2005}
\accepted{11 Nov 2005}
\publonline{later}


\abstract{%
We study the behaviour of the van der Pol oscillator when
either its damping parameter $\mu$ or its nonlinearity parameter $\xi$
is subject to additive or multiplicative random noise. Assuming
various power law exponents for the relation between the oscillating
variable and the sunspot number, for each case we map the parameter
plane defined by the amplitude and the correlation time of the
perturbation and mark the parameter regime where the sunspot number
displays solar-like behaviour.    Solar-like behaviour is defined here
as a good correlation between the rise rate and cycle amplitude {\it
and} the lack of a good correlation between the decay rate and
amplitude, together with significant ($\ga 10$\,\%) r.m.s. variation in
cycle lengths and cycle amplitudes. It is found that perturbing $\mu$
alone the perturbed van der Pol oscillator does not show  solar-like
behaviour. When the perturbed variable is $\xi$, solar-like behaviour is
displayed for perturbations with a correlation time of about 3--4 years
and significant amplitude. Such studies may provide useful constraints
on solar dynamo models and their parameters. } \maketitle

\section{Introduction}

A complete spatial truncation of the nonlinear $\alpha\Omega$ dynamo
equations with a dimensional analysis of some of the terms is known to
give rise to a nonlinear oscillator equation of the form
\begin{equation}
\label{nemlinoegy}
\ddot B = -\omega^2 B - \mu(\xi B^2-1)\dot{B} -\gamma B^3
\end{equation}
where $B$ is the amplitude of the toroidal magnetic field, and the
parameters $\mu$, $\xi$ and $\gamma$ may be expressed by the dynamo
parameters (dynamo number, meridional flow amplitude, nonlinearity
parameters) \citep{mininni2001,lopes2011}. This is a combination of the
van der Pol and Duffing oscillators, the two most widely studied
nonlinear oscillator problems.

In the past decade, the possibility of representing the sunspot number
series with such a nonlinear oscillator model was explored in a number
of papers (e.g. \citealt{mininni2000}; \citealt{pontieri2003};
\citealt{lopes2009meridional}; \citealt{passos2012}). These studies
demonstrated that with a suitable choice of parameters, the overall
phase space structure of the sunspot number series can be well
reproduced and cycle-to-cycle variations may also be qualitatively well
modeled by admitting a stochastic perturbation to one or more of the
model parameters.

A more quantitative study of this problem should, however, also examine
whether the behaviour of a stochastically perturbed oscillator of the
form (\ref{nemlinoegy}) adheres to the known regularities in the
cycle-to-cycle variation of solar activity. The most important such
regularity is the Waldmeier effect. In its currently adopted formulation
the effect consists in a good correlation (coefficient $r\simeq 0.85$)
between the rise rate of a cycle and its maximum amplitude
(\citealt{lantos}; \citealt{cameron}). On the other hand the {\it lack}
of a statistically significant correlation between the decay rate and
the cycle amplitude forms an equally important quantitative constraint.
Taking into account the length of the sunspot number series this implies
that $|r|\la 0.5$ for any such correlation.

In order to study this problem we have started a systematic
investigation of the parameter space of stochastically perturbed
nonlinear oscillators of the type (\ref{nemlinoegy}). As a first step,
here we consider a pure van der Pol oscillator, neglecting the last term
in equation (\ref{nemlinoegy}).

In the next section we present our stochastically perturbed van der Pol
oscillator model in detail while results of the Monte Carlo simulations
are presented and discussed in Section \ref{results}. Section 4
concludes the paper.

\section{Objectives and Method}

    \subsection{Objectives}\label{requirements}
    The equation of the van der Pol oscillator used in the present study is
    \begin{equation}
    \label{vdpealap}
    \ddot x=-\omega^2x-\mu\left( \xi x^2-1\right)\dot{x}.
    \end{equation}
    The parameters in the equation are: frequency $\omega$, dumping
    parameter $\mu$, and nonlinearity $\xi$. Of these, $\omega$ and
    $\xi$ determine the cycle length and amplitude, respectively, while
    $\mu$ determines the degree of asymmetry in cycle profiles.

    We assume that the oscillator variable $x$ is proportional to the
    amplitude of the toroidal magnetic field and in turn it is related
    to the sunspot number as a power law $SSN=x^n$. For $n$ the values
    1, 1.5 and 2, often used in the literature (e.g.
    \citealt{bracewell1988}; \citealt{mininni2000}), were used in
    various cases.

    \citet{mininni2000} fitted this oscillator to the entire sunspot
number series registered between 1816 and 2000, assuming $n=2$. In our
simulations we arbitrarily chose these fits as the
unperturbed values of the parameters:
    $\omega_0=0.2993, \mu_0=0.2044, \xi_0=0.0102$.
    (Note that this choice is clearly arbitrary for $n\neq 2$ and, due to
    our definition of $\xi$ being different from that of Mininni et al.,
    even for $n=2$. However, different choices would only result in
    different cycle lengths and amplitudes which may be accounted for by
    proper normalization.)

Previous authors have found that adding stochastic perturbations to the
oscillator parameters, the phase space structure of the solar activity
cycle can be reproduced qualitatively well. However, it has not been
studied whether and under what conditions such stochastically perturbed
oscillator models give rise to cycle variations that compare favorably
with the variations in the  observed solar cycle also in a quantitative
manner. The single most important observed quantitative constraint on
solar cycle variations is the Waldmeier rule; in its current formulation
(\citealt{cameron}) this consists in a strong ($r\ga 0.8$) positive
correlation between the rise rate of solar cycles and their maximal
amplitudes. However, an equally important constraint is the {\it lack}
of a statistically significant correlation between the decay rate of
cycles and their amplitudes; taking into account the number of cycles on
which this conclusion is based, this implies that any such correlation
should have $|r|\la 0.5$. Finally, we know that there is a significant
($\sim 20$--$30\,$\%) cycle to cycle variation in both cycle amplitudes
and cycle lengths.

In summary, for a perturbed oscillator to qualify as solar-like, it is
required to adhere to the following criteria. (1)
$r_{\mathrm{rise}}>0.8$; (2) $|r_{\mathrm{decay}}|<0.5$; (3) r.m.s.
variation of cycle amplitudes (rms$_A$) and periods (rms$_T$) should
exceed 10\,\%

    \subsection{Method}

For a systematic study of the effect of various types of perturbations
of the oscillator parameters on the resulting sunspot number series, a
large number of Monte Carlo simulations were performed. Each of these
simulations consisted in a numerical integration of equation (2) with
random noise acting on one of its parameters.

For stochastic fluctuations acting on model parameters two cases were
considered.
One was additive noise:
    \begin{equation}
    \label{additivedef}
    \mu(t)=\mu_0+d(t) \quad\mathrm{ or }\quad \xi(t)=\xi_0+d(t)
    \end{equation}
    and the other was multiplicative noise:
    \begin{equation}
    \mu(t)=\mu_0\mathrm{e}^{d(t)}\quad\mathrm{ or }\quad
    \xi(t)=\xi_0\mathrm{e}^{d(t)}.
    \label{multipldef}
       \end{equation}
Here the noise $d(t)$ was modeled as a Gaussian random process. For the
time dependence of $d(t)$ two different approaches were applied again.
In one case, $d(t)$ is a piecewise constant function keeping a constant
value for a time $T_{corr}$, then taking another random value $R$.  In
the other case, in order to produce a more realistic, continuously
varying noise,  $d(t)$ was modelled as an Ornstein--Uhlenbeck process
(\citealt{gillespie1996}, \citealt{longtin2010}) according to
\begin{equation}
d(t)=d(t-dt)-K d(t-dt)dt + \Delta R\sqrt{dt}
\end{equation}
In this case the inverse of the relaxation constant $K$ plays the role
of a time constant, taking the place of $T_{corr}$.  In both cases, the
random value $R$ is taken out of a Gaussian distribution of half width 1
and the amplitude of the perturbation is set by the parameter
$\Delta$.

The effects of time dependence in $\mu$ and $\xi$ were considered separately,
i.e. the equation used was either
   \begin{equation}
    \label{vdp-mu}
    \ddot x=-\omega_0^2x-\mu(t)\left( \xi_0 x^2-1\right)\dot{x},\\
    \end{equation}
or
    \begin{equation}
    \label{vdp-xi}
    \ddot x=-\omega_0^2x-\mu_0\left[ \xi(t) x^2-1\right]\dot{x}.
    \end{equation}

For both cases, four combinations of perturbation types are possible:
additive or multiplicative and piecewise constant or continuous.
Combined with the 3 possible values of $n$ mentioned above, this gives
rise to 24 model families. In each family the parameter plane defined
by the  amplitude and timescale of the perturbation
($\Delta$--$T_{corr}$ plane or $\Delta$--$K$ plane) was covered by a
grid of models (order of a hundred models in each family). The part of
the parameter plane covered corresponds to correlation times of 1 to
20 years and perturbation amplitudes from 0.03 to 0.67. Occasionally,
when the randomly varying perturbation leads to excessive values of
$x$, the simulations would become unstable. To avoid this, further
restrictions on the values of the fluctuationg parameters were
introduced. In the additive case, the value of the perturbed parameter
was not allowed to exceed twice its unperturbed value, while in the
multiplicative case it was restricted to be within one order of
magnitude from the unperturbed value. In order to obtain statistically
meaningful results, in each of the cases considered the oscillator
equation was integrated for 2000 years.

    \begin{figure}
    {\includegraphics[width=\linewidth]{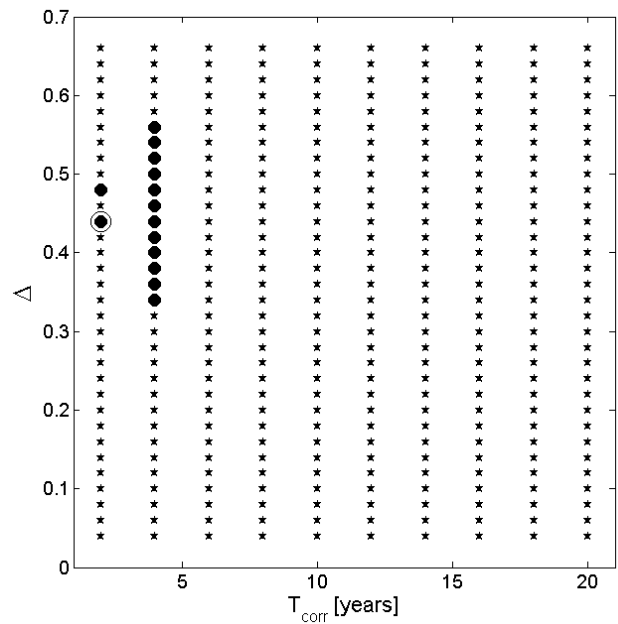}
    \caption{Our model grid (small asterisks) on the parameter plane in
the case of a piecewise constant perturbation applied multiplicatively
to $\xi$, with $n=2$. Larger dots ($\bullet$) indicate the runs
displaying solar-like behaviour as defined in the text
(Sect.~\ref{requirements}). The encircled ``representative'' model has
$\Delta=0.44$, $T_{corr}=2$ years.}
    \label{paramtart}}
    \end{figure}

\section{Results and discussion}\label{results}

Having generated the model grids as described above, adherence to  the
criteria set of solar-like behaviour (see end of Sect.
\ref{requirements}) was checked for each model. The following results
were obtained.

    \subsection{The perturbed parameter is $\mu$}
    If the perturbed parameter is $\mu$, the model could only
approximate the expected behaviour. The van der Pol oscillator (Eq.
\ref{vdp-mu}) reproduced the Waldmeier effect with several input
parameter pairs ($\Delta$--$K$ and $\Delta$--$T_{corr}$); however, none
of the models did simultaneously satisfy all the criteria of solar-like
behaviour.

    \subsection{The perturbed parameter is $\xi$}

    If the time dependent parameter is $\xi$, the model (Eq.
\ref{vdp-xi}) shows all the expected features, detailed in Section
\ref{requirements}, in several cases. Optimal results were found in the
case of multiplicative perturbations. The form of the time dependence of
the perturbation (piecewise constant or coninuous) made no major
difference to the results. As an example, in Figure~1 we present
the map of the parameter plane for the case of a piecewise constant,
multiplicative perturbation with $n=2$.

\begin{figure*}[t]
    {\includegraphics[width=\textwidth,height=0.4\textwidth]{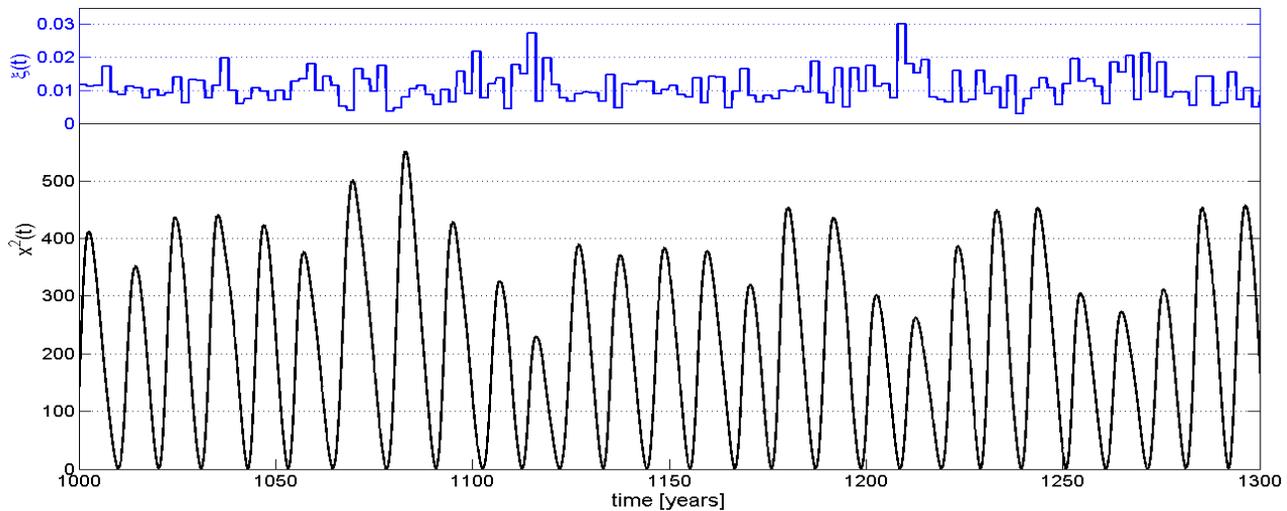}
    \caption{A 300 year long segment of the ``representative'' model
(encircled in Fig.~1) as an example for the behaviour of a
stochastically perturbed van der Pol oscillator displaying solar-like
behaviour. Shown are the temporal variations of the oscillator parameter
$\xi$ (top), and of the sunspot number (bottom), assumed to scale with
$x^{2}(t)$ in this case. The properties of the model are summarized in
Table \ref{osszehtabla}. }
    \label{xifm-x2}}
    \end{figure*}

    The Waldmeier effect is displayed by all the models in the  grid
presented in Figure~1. However, all criteria of solar-like behaviour are
only satisfied in a minority of cases, with correlation times of $\sim
2$--$4$ years and an appropriately chosen perturbation amplitudes
($\sim\pm 50\,$\%). The cycles displayed by one such model  are
presented in Figure \ref{xifm-x2} while Table \ref{osszehtabla} presents
the attributes of the model compared to the observed characteristics of
the solar cycle. It is apparent from the figure that episodes of
sustained positive fluctuations in $\xi$, for example after $t=1100$,
lead to solar cycles of a reduced amplitude. Similarly, stronger cycles
are caused by episodes of  sustained negative fluctuations in $\xi$
values, for instance in the period between
$t=1051\mathrm{\;and\;}1100$.

We note that the observed sunspot number series seems to display
sustained periods of low or high activity lasting for up to three to six
cycles. This seems to indicate a memory effect extending over several
solar cycles; however, evidence for this is still controversial, so it
was not included among the criteria for solar-like behaviour. In any
case, such behaviour is not seen in our models.

The occasional occurrence of extended grand minima like the Maunder
minimum is also an empirical characteristic of solar activity. Again,
this was not included among our criteria as reproducing grand minima was
beyond the scope of the present study which focuses on reproducing the
``normal'' state of solar activity, outside grand minima, as evidenced
in the official sunspot number series. Nevertheless, a little
experimentation showed that grand minima may be obtained in our models
with a time dependent $\xi$ if the restrictions on the values of
perturbed parameter were relaxed and $\xi(t)$ was allowed to reach
higher values ($\xi(t)<50\xi_0$). However, with this new limit, the
model either became unstable at some point or, in addition to the grand
minima, extremely long ``supercycles'' of $\sim 100$ years also
appeared.

    \begin{table*}
    \centering
    \caption{Comparison of the observed attributes of the solar sycle
and of the van der Pol oscillator model, shown in Figure \ref{xifm-x2}. The
attributes were obtained by analyzing the entire 2000 year long $x^2(t)$
function of the oscillator.}
    \begin{tabular}{r c c c c c c }
    & $r_{\mathrm{rise}}$ & $r_{\mathrm{decay}}$ & Rise time & Period & rms$_T$ & rms$_A$ \\
    &  &  &(year) & (year) & (\%) &  (\%)\\
    \hline     \hline
    Solar cycle & 0.85 & 0.35 & 3.5 & 11.02 & 11 & 30\\
    van der Pol  & 0.84 & -0.49 & 4.2 & 10.92 & 10 & 20\\
    \end{tabular}
    \label{osszehtabla}
    \end{table*}

The amplitude and cycle length distributions of the models  were also
studied by performing a run for a 11000 year long interval for one
representative case in each model family. For models that were stable
and lacked the extremely long supercycles mentioned above, the histogram
of cycle amplitudes was close to Gaussian, without an excess at low
amplitudes. (Such an excess is the hallmark of grand minima as a separate
state of activity, cf. \citealt{usoskin}.)

The histogram of cycle lengths shows a maximum at $\sim 11$ years, the
total range going from $\sim 8$ years up to $\sim20$ years. Such
unusually long periods occur once or twice during the 11000 year long
time interval.

\section{Conclusion}

The present study is only the first step in a more extensive systematic
investigation of the parameter space of the stochastically perturbed
nonlinear oscillators potentially representing the sunspot number series.

Here we focused on the behaviour of stochastically perturbed van der Pol
oscillator when only one of its parameters (either the damping $\mu$ or
the nonlinearity $\xi$) is varied. We focused on reproducing the
Waldmeier effect and other known features of the solar cycle. It was
found that in certain parameter ranges the stochastically perturbed van
der Pol oscillator does display all the expected characteristics
simultaneously but only when the perturbed oscillator parameter is
$\xi$.

Another interesting conclusion from these studies is that a good
correlation between cycle amplitude and rise rate (Waldmeier effect) is
displayed by a fairly wide class of nonlinear oscillators; at the same
time the {\it lack} of a similar good correlation between cycle
amplitudes and decay rates is a potentially even more stringent
condition for solar-like behaviour that is in fact often harder to
reproduce than the Waldmeier effect itself.

Our finding  that random perturbations of $\mu$ alone do not lead to
solar-like behaviour is of some interest as in at least one earlier
study (\citealt{lopes2008}) cycle variations in the observed SSN series
were interpreted in terms of variations in the speed of meridional
circulation which, in the truncated flux transport dynamo considered
there, mainly influence the oscillator through the damping parameter.
This underlines that while fitting the observed cycles with a nonlinear
oscillator with varying parameters results in parameter variations that
may seem to be random at first sight, the random nature of such
variations is actually not demonstrated and a systematic study of
an ensemble of oscillators with truly random parameter variations is
needed to confirm or discard solar-like behaviour.

It should be stressed, however, that this conclusion is still of a
preliminary nature and it may change if e.g. the perturbations in
$\mu$ and $\xi$ are assumed to be interrelated, as expected for a
flux transport dynamo.

Possibilities for future extensions of this study therefore include a
joint perturbation of $\mu$ and $\xi$ and the inclusion of a (possibly
also perturbed) third order term in the oscillator equation, as in
equation (1).

\acknowledgements
This research was supported by the Hungarian Science Research Fund (OTKA grants
no.\ K83133 and 81421).

\newpage

\end{document}